\documentclass[journal]{IEEEtran}

\usepackage{graphicx}
\graphicspath{{fig/}}
\usepackage{cite}
\usepackage{amsmath,amssymb,amsfonts}
\usepackage{algorithmic}
\usepackage{graphicx}
\usepackage{textcomp}
\usepackage{subfigure}
\usepackage[ruled]{algorithm2e}
\usepackage{multirow}
\usepackage{tabularx}
\usepackage{bm}
\usepackage{stfloats}
\usepackage{diagbox}
\usepackage{slashbox}
\usepackage{flushend}
\usepackage{balance}
\usepackage{soul}
\usepackage{float}

\soulregister\cite7
\soulregister\citep7
\soulregister\citet7
\soulregister\ref7
\soulregister\pageref7

\usepackage{xcolor,soul,framed} 

\colorlet{shadecolor}{yellow}

\begin{document}
	\title{UAV Trajectory Planning in Wireless Sensor Networks for Energy Consumption Minimization by Deep Reinforcement Learning}	
	\author{Botao~Zhu,~\IEEEmembership{}
		Ebrahim~Bedeer,~\IEEEmembership{Member,~IEEE,}
		~Ha H. Nguyen,~\IEEEmembership{Senior Member,~IEEE,} Robert~Barton,~\IEEEmembership{Member,~IEEE}, and Jerome Henry,~\IEEEmembership{Senior Member,~IEEE}
		\thanks{B. Zhu, E. Bedeer, and H. H. Nguyen are with the Department of Electrical and Computer Engineering, University of Saskatchewan, Saskatoon, Canada S7N5A9. Emails: \{botao.zhu, e.bedeer, ha.nguyen\}@usask.ca.}
		\thanks{R. Barton and J. Henry are with Cisco Systems Inc. Emails: \{robbarto, jerhenry\}@cisco.com.} 
		\thanks{Copyright (c) 2015 IEEE. Personal use of this material is permitted. However, permission to use this material for any other purposes must be obtained from the IEEE by sending a request to pubs-permissions@ieee.org.}
	}

	\maketitle
	\begin{abstract}
		Unmanned aerial vehicles (UAVs) have emerged as a promising candidate solution for data collection of large-scale wireless sensor networks (WSNs). In this paper, we investigate a UAV-aided WSN, where cluster heads (CHs) receive data from their member nodes, and a UAV is dispatched to collect data from CHs along the planned trajectory. We aim to minimize the total energy consumption of the UAV-WSN system in a complete round of data collection. Toward this end, we formulate the energy consumption minimization problem as a constrained combinatorial optimization problem by jointly selecting CHs from nodes within clusters and planning the UAV's visiting order to the selected CHs. The formulated energy consumption minimization problem is NP-hard, and hence, hard to solve optimally. In order to tackle this challenge, we propose a novel deep reinforcement learning (DRL) technique, pointer network-A* (Ptr-A*), which can efficiently learn from experiences the UAV trajectory policy for minimizing the energy consumption. The UAV's start point and the WSN with a set of pre-determined clusters are fed into the Ptr-A*, and the Ptr-A* outputs a group of CHs and the visiting order to these CHs, i.e., the UAV's trajectory. The parameters of the Ptr-A* are trained on small-scale clusters problem instances for faster training by using the actor-critic algorithm in an unsupervised manner. At inference, three search strategies are also proposed to improve the quality of solutions. Simulation results show that the trained models based on 20-clusters and 40-clusters have a good generalization ability to solve the UAV's trajectory planning problem in WSNs with different numbers of clusters, without the need to retrain the models. Furthermore, the results show that our proposed DRL algorithm outperforms two baseline techniques.
	\end{abstract}
	
		
	\section{Introduction}\label{SecI}
	\label{sec:introduction}
	\IEEEPARstart{T}{he} use of unmanned aerial vehicles (UAVs) has recently attracted a lot of attention from both the research community and industry. UAVs have been used for a variety of purposes \cite{M. Mozaffari}, such as environmental monitoring, mobile cloud computing, disaster management, security operations, and wireless power transfer, to name a few. The popularity and widespread applications of UAVs are due to their many advantages, such as cost-effectiveness, having line-of-sight (LoS) links with the ground devices, mobility, and reliable network access \cite{Y. Zeng 2016}.
		
	In order to overcome the limitations of the tradition wireless sensor networks (WSNs), extensive research has been done on the integration of UAVs and WSNs for long-distance mission communications where UAVs are considered as mobile sinks for receiving data from cluster heads (CHs), and then, they can transmit the collected data to terrestrial BSs for further processing. Using UAVs as mobile sinks can reduce the energy consumption of ground nodes in WSNs when compared to the traditional multi-hop WSNs which transmit data from each node to the sink node over a long distance or several hops \cite{P. Nayak}. Despite the advantages, the integration of UAVs and WSNs still grapples with many challenges. First of all, due to the limited energy source carried by UAVs, the service range of UAVs is constrained by the reality that they cannot travel very long distances or fly for long periods of time. Second, battery life of sensor nodes in WSNs is typically limited, and in many cases it is hard to regularly replace their batteries. As a result, frequent communication with UAVs can cause sensor nodes to exhaust their energy rapidly. Hence, it is important to study the energy saving problem in UAV-enabled WSNs.

	\subsection{Motivation}
	Prior works on the energy consumption minimization for UAV-enabled WSNs can be classified into three categories depending on the objectives. The first category only considers minimizing the UAV's energy consumption, e.g., \cite{Q. Song, Y. Zeng}. In contrast, the second category considers only minimizing the energy consumption of the ground devices in the UAV-aided wireless networks, e.g., \cite{C. Zhan and H. Lai, J. Baek}. In the third category, the energy of both the UAV and the ground devices are taken into account when minimizing the energy consumption of the UAV-enabled system, e.g., \cite{M. B. Ghorbel, D. Yang}. However, most of the aforementioned studies assume the UAV directly communicates with each device of the ground wireless network. In this case, if the UAV flies over all devices in a large-scale WSN, it would lead to a long flight trajectory for the UAV which increase its energy consumption. As a result, the UAV may run out of its energy in flight or may need to recharge its battery frequently.
	
	Motivated by the aforementioned works, in this paper we investigate the problem of minimizing the total energy consumption of the UAV and the ground devices in a clustered WSN, which has not been well researched in prior works. We assume that devices on the  ground have been clustered according to some specific criterion, e.g., based on their geographical locations; hence, clustering techniques for WSNs will not be discussed in this paper. In each pre-determined cluster in our system model, one of the ground devices will be selected as the CH, which is responsible for collecting data from the non-CH devices in the same cluster. Hence, the UAV only needs to visit a set of CHs for gathering data along the planned trajectory that is determined by locations of the ground CHs. The selection of CHs affects both the energy consumption of the ground devices and the UAV, as shown in the following illustrative example. 

\begin{figure}[t]
		\centering
		\includegraphics[width=1\linewidth]{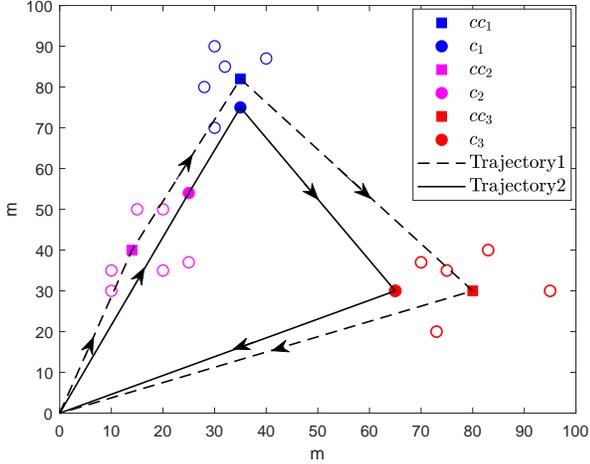}
		\caption{Comparison of different UAV's trajectories.}
		\label{figure1}
	\end{figure}

Consider the case where nodes of three clusters are deployed in a given area as shown in Fig.~\ref{figure1}. There are two candidate solutions of potential CHs. One possible solution, i.e., trajectory 1, selects the ``center'' node of each cluster as the CH, such as $cc_1, cc_2, cc_3$, and the other possible solution, i.e., trajectory 2, selects non-center nodes as the CHs, e.g. $c_1, c_2, c_3$.	The start/end location of the UAV is (0, 0). If the UAV chooses to follow trajectory 1, the energy consumption between CHs and their member nodes will be minimal because the Euclidean distances between member nodes and their CHs are minimal (on average) per cluster \cite{B. Zhu}. However, this will lead to an increase in energy consumption of the UAV as its flight trajectory may be longer, and hence, may not be optimal from the start point to the end point, which can be seen from trajectory 1 in Fig. 1. On the contrary, if the UAV goes through trajectory 2, it will consume less energy in flight because trajectory 2 is shorter than trajectory 1. But, the communication energy consumption between CHs and their member nodes will be higher in this case. From the above simple example and discussion, it is clearly important and relevant to study the energy-efficient UAV's trajectory planning in clustered WSNs in order to minimize the overall energy consumption of the UAV and the ground network.

	\subsection{Related Works}
	\subsubsection{UAV Trajectory Planning}
	
Energy-efficient trajectory planning for UAVs has recently attracted significant research interest, and multiple solutions have been proposed for UAV-enabled wireless networks. In general, existing solutions for energy-efficient UAV trajectory planning can be loosely classified into two categories: non-machine learning-based methods and machine learning-based methods. In the first category, researchers mostly use mathematical programming or heuristic algorithms to solve the trajectory optimization problem. However, the computation time of mathematical programming algorithms may increase exponentially as the problem size increases, e.g., \cite{M. Samir, S. Zhang}. Although some heuristic algorithms are applied to design the energy-efficient path in the UAV-enabled wireless networks, such as ant colony optimization \cite{A. A. Al-Habob} and cuckoo search \cite{K. Zhu}, they usually cannot fully adapt to the increasing complexity of scalable wireless networks.

Regarding the machine learning-based category, deep reinforcement learning (DRL) and reinforcement learning (RL) are the most common techniques in solving the UAV's trajectory planning. In \cite{C. H. Liu}, the authors propose a DRL-based method which is composed of two deep neural networks (DNNs) and deep deterministic policy gradient (DDPG) to maximize the energy efficiency for a group of UAVs by jointly considering communications coverage, energy consumption, and connectivity.  In order to minimize the UAV's transmission and hovering energy, the authors in \cite{Y. Yuan} formulate the energy-efficient optimization problem as a Markov decision process. Then, they use two DNNs and the actor-critic-based RL algorithm to develop an online DRL algorithm that shows a good performance in terms of energy savings. In \cite{R. Ding}, the authors jointly optimize the UAV's 3D trajectory and the frequency band allocation of ground users by considering the UAV's energy consumption and the fairness of the ground users.  A DDPG-based DRL algorithm is developed to generate the energy-efficient trajectory with fair communication service to ground users. In \cite{B. Zhang}, with the aim of designing an energy-efficient UAV's route for long-distance sensing tasks, the authors propose a DRL-based framework where convolutional neural networks (CNNs) are used for extracting features and the deep Q-network (DQN) is utilized to make decisions. Towards realizing green UAV-enabled Internet of Things (IoT), the authors in \cite{W. Liu} formulate the UAV's path planning problem as a dynamic decision optimization problem, which is solved by dueling DQN. The aforementioned machine learning-based methods show strong ability to handle complex wireless environments and effectively learn the UAV's trajectory policy from experiences;  however, they are implemented by using some common neural network models, such as DNNs, DQN, and CNNs. Instead of common neural network models, in this paper we exploit the appealing concept of sequence-to-sequence learning that originally emerged in \emph{Natural Language Processing} to design the DRL algorithm to solve the UAV's path planning problem in clustered WSNs.
	
	 \subsubsection{Sequence-to-Sequence Learning}
	 Sequence-to-sequence learning has shown great success in machine translation where sentences are mapped to correct translations \cite{H. Zhang}. Over the last several years, many neural networks based on sequence-to-sequence models have been proposed in different applications. For example, pointer network is one of the extensively studied models because of its excellent ability in solving sequence decision problems. In \cite{O. Vinyals}, pointer network is trained in a supervised fashion to solve the traveling salesman problem (TSP). The work of \cite{I. Bello} uses a RL-based unsupervised method to train the pointer network and obtains better results when compared to the supervised learning in \cite{O. Vinyals}. In \cite{J. J. Q. Yu}, the authors propose a structural graph embedded pointer network to develop online vehicular routes in intelligent transportation systems. The authors in \cite{Z. Li} propose a modified pointer network to solve the keyword recommendation problem in sponsored search advertising system. In \cite{M. Nazari}, a simplified pointer network is introduced to solve Vehicle Routing Problem (VRP) in dynamic traffic environments. Different from the above-discussed research works, we extend the state-of-the-art pointer network-based DRL to solve the UAV's trajectory planning problem. Our contributions are elaborated in the next subsection.
		
	\subsection{Contributions}

	We aim to minimize the overall UAV-WSN's energy consumption by designing an efficient UAV's trajectory in a clustered WSN. Since the UAV's visiting order to the ground CHs can be seen as a sequence decision problem, we propose a sequence-to-sequence learning-based DRL strategy with pointer network to deal with the challenging trajectory planning problem. The pointer network can capture the relation between a problem instance and its solution by using a sequence-to-sequence neural network. It has been demonstrated to be an effective method to solve some NP-hard problems, such as TSP \cite{I. Bello} and VRP \cite{M. Nazari}. Hence, it is expected that the pointer network-based DRL algorithm is also promising for solving the problem of the UAV trajectory planning for the UAV-WSN system. The main contributions of this paper are summarized as follows:
	
	\begin{enumerate}
		\item We consider a UAV-enabled energy-efficient data collection framework for clustered WSNs. We formulate an optimization problem to minimize the energy consumption of the entire UAV-WSN system by jointly designing the UAV's trajectory and selecting CHs in pre-determined clusters of the ground WSN.
				
		\item We show that the UAV's trajectory planning problem in the clustered WSN can be seen as a sequence of decisions. Hence, a sequence-to-sequence pointer network-A* (Ptr-A*) model is proposed to solve the formulated problem. Particularly, the  pointer network is utilized to model the visiting order of all ground clusters, and A* algorithm\cite{V. Razo and H. Jacobsen} is used to efficiently select CHs from clusters' ground nodes. The UAV's start point and all clusters, as the input, are fed into the Ptr-A* model, and its output is a set of CHs and the visiting order to these CHs, i.e., the UAV's trajectory.
		
		\item We use a self-driven learning mechanism that only needs the reward calculation to train the parameters of Ptr-A* network on problem instances with small-scale clusters for faster training.

		\item Our proposed DRL method has an excellent generalization capability with respect to the number of clusters used for training. In other words, given a new problem instance with any number of clusters, the trained model can automatically generate a trajectory for the UAV to visit clusters, without retraining the new model.
		
		\item We perform extensive simulations to demonstrate that the proposed DRL method outperforms other baseline techniques when considering both the computation times and energy consumption results.
	\end{enumerate}
	
	The rest of this paper is organized as follows. Section \ref{SecIII} presents the system model and the problem formulation. Section \ref{SecIV} describes the proposed DRL algorithm. Section \ref{SecVI} provides simulation results. Finally, Section \ref{SecVII} concludes the paper.

	\section{System Model and Problem Formulation}\label{SecIII}
	
	As mentioned earlier, we assume that devices on the ground have been clustered according to some specific criterion, e.g., based on their geographical locations; hence, clustering techniques for the ground WSN are not discussed in this work.
	In particular, we consider $K$ clusters of sensor nodes  $\{\bm{G}_1,\dots, \bm{G}_K\}$ located in the sensing (service) area for data collection. Each cluster contains $N$ nodes, one of which is the CH, represented by $b^{}_k\in \bm{G}_k$, that will be selected by our proposed algorithm. We assume that only one rotary-wing UAV is dispatched to visit CHs to collect data from the ground network. The UAV takes off from the start position $b_0$ and then back to $b_0$ after finishing the data collection task. The trajectory of the UAV should contain the start/end hovering position $c_0$ corresponding to $b_0$, and the $K$ target hovering positions $\{c_1,\dots, c_K\}$, which are vertically above the ground CHs. Hence, the trajectory planning problem of the UAV can be seen as a permutation of $(K+1)$ hovering positions. It is obvious that the locations of CHs determine the flight trajectory, and hence, the energy consumption of UAV and the ground nodes. We consider a three-dimensional (3D) Cartesian coordinates system to define the positions of ground nodes and the UAV. The position of the CH of the $k$-th cluster is $b_k = \left(x_k, y_k, 0\right)$. Correspondingly, the coordinate of the UAV's hovering position $c_k$ can be represented by $\left(x_k, y_k, H\right)$, where $H$ is the fixed flight height of the UAV. Similarity, the coordinate of the $n$-th member node of the $k$-th cluster $b_k^{(n)}$ is denoted as $\left(x_k^{(n)}, y_k^{(n)}, 0\right), n=1,\dots,N-1$, $k=1,\dots,K$, and $b_k^{(n)} \neq b_k$.

	\subsection{Channel Model}
	There is a number of channel models that have been developed for UAV communications, e.g., \cite{Z. Ma, Z. Lian}. In this work, we consider a simple air-to-ground channel model that is described as follows.
	For ground-to-air communication, there is a certain probability that each CH $b_k$ has a LoS view towards the UAV when it hovers at the hovering position $c_k$. This probability typically depends on the environment and elevation angle, and is given by \cite{M. B. Ghorbel}
	\begin{equation}
	    P_{\text{LoS}} = \frac{1}{1+\eta\exp{\left(-\beta[\tau-\eta]\right)}},
	\end{equation}
	where $\eta$ and $\beta$ are constants determined by environment, and $\tau=\frac{180}{\pi}\times\sin^{-1}\left({\frac{H}{d_k}}\right)$, where $d_k$ is the distance between $b_k$ and $c_k$. Since it is assumed that each hovering position is directly above the corresponding CH, one has $d_k=H$. Obviously, the non-line-of-sight (NLoS) probability is given by $P_{\text{NLoS}} = 1-P_{\text{LoS}}$. The average path loss between each CH and the UAV can be expressed as \cite{M. B. Ghorbel}
	\begin{equation}
	    \overline{P}_{\text{loss}} = P_{\text{LoS}} \left(K_0 + \mu_{\text{LoS}} \right) + P_{\text{NLoS}}\left(K_0 + \mu_{\text{NLoS}} \right)
	\end{equation}
	where $\mu_{\text{LoS}}$ and $\mu_{\text{NLoS}}$ are the mean values of the excessive path losses in LoS and NLoS links, respectively, $K_0 = 10\alpha \log_{10}\left(\frac{4\pi f_c H}{c}\right)$, $\alpha$ is the path loss exponent, $c$\ is the speed of light, and $f_c$ is the carrier frequency. Thus, the average data rate from each CH to the UAV can be computed as \cite{M. B. Ghorbel}
	\begin{equation}
	    r_{\text{data}} = B_{\text{width}}\log_2\left(1+\frac{P_{\text{CH}}}{\overline{P}_{\text{loss}}N_{0}}\right)
	\end{equation}
	where $ B_{\text{width}}$ is the available bandwidth, $N_0$ is the noise power spectral density, and $P_{\text{CH}}$ is the transmit power of each CH.
		
	\subsection{UAV's Energy and Trajectory Model}
	We assume that the UAV supports a flying-hovering mode without considering acceleration-deceleration patterns.  After the UAV flies to the hovering position $c_k$ with a fixed speed $v_{\text{UAV}}$, it hovers there and transfers a beacon frame to wake up the corresponding CH $b_k$ from sleep mode to active model. Then, $b_k$ starts to collect data from its member nodes by time-division multiple access (TDMA) and forwards the collected data to the UAV. At each hovering position, the energy consumption of the UAV includes two parts: communication-related energy and hover-related energy. The hovering power is given by \cite{M. B. Ghorbel}, \cite{D. Hulens}
	\begin{equation}
	    P_{\text{hover}} = \sqrt{\frac{\left(m_{\text{tot}}g\right)^3}{2 \pi r^2_{p}n_{p}\rho}}
	\end{equation}
	where $g$ is the earth gravity, $\rho$ is the air density, $n_p$ is the number of propellers, $r_p$ is the propeller radius, and finally $m_{\text{tot}}$ is the mass of the UAV. Thus, the energy consumed by the UAV at each hovering position $c_k$ is given by
	\begin{align}
	    E_{c_k} & = T_{k}(P_{\text{hover}} + P_{\text{com}}) \nonumber \\
	    & = \frac{D_k}{r_{\text{data}}}(P_{\text{hover}} + P_{\text{com}})
	\end{align}
	where $T_k$ is the total hovering time of the UAV at $c_k$, $D_k$ is the amount of data that needs to be transferred from CH $b_k$ to the UAV, and $P_{\text{com}}$ is the communication power of the UAV. In order to simplify the analysis, we assume that the hovering time is equal to the data transmission time from $b_k$ to the UAV.
	
	The horizontal movement power is assumed as a linear function of the UAV's flight speed $v_{\text{UAV}}$, which is expressed as\cite{M. B. Ghorbel}, \cite{D. Hulens}
	\begin{equation}
	    P_{\text{move}} = \frac{P_{\text{max}}-P_{\text{idle}}}{v_{\text{max}}}v_{\text{UAV}} + P_{\text{idle}}
	\end{equation}
	where $v_{\text{max}}$ is the maximum speed of the UAV, $P_{\text{max}}$ and $P_{\text{idle}}$ are the hardware power levels when the UAV is moving at full speed and when the UAV is in idle state, respectively. Because the UAV needs to start from the start hovering location $c_0$, goes through all target hovering positions $c_1,\dots, c_K$, and then back to $c_{0}$, the total energy consumption of the UAV in flight is given by\cite{M. B. Ghorbel}, \cite{D. Hulens}
    \begin{eqnarray}
            E_{\text{flight}} = T_{\text{flight}}\left(P_{\text{hover}} + P_{\text{move}}\right)
    \end{eqnarray}
    where $T_{\text{flight}}$ is the total flight time, which can be expressed as
    \begin{equation}
        T_{\text{flight}} = \frac{1}{v_{\text{UAV}}}\sum_{i=0}^{K} \sum_{\substack{j=0 \\ j \neq i}}^{K} d_{c_{i},c_{j}}L_{c_{i},c_{j}},\quad \forall{c_i,c_j} \in {\bm{C}}
    \end{equation}
	where $\bm{C} = \{c_0,c_1,\dots, c_K\}$, $c_k$ is determined by $b_k$, $b_k \in \bm{G}_k$, and  $L_{c_i,c_j}$ specifies whether the UAV travels from stop position $c_i$ to $c_j$, which is defined as
	\begin{equation}
	\label{visit}
	L_{c_i,c_j} = \left\{ \begin{array}{ll}
	1,& \text{if the path goes from $c_i$ to $c_j$}\\
	0,& \text{otherwise}.
	\end{array} \right.
	\end{equation}
	The quantity $d_{c_i,c_j}$ is the Euclidean distance between $c_i$ and $c_j$, which is given by
	\begin{equation}
	d_{c_i,c_j} = ||c_i-c_j|| = ||b_i-b_j||.
	\end{equation}
	In order to meet the requirements of the UAV's trajectory, we need to consider the following constraints:
	\begin{equation}
	\label{in}
	\sum_{\substack{i=0 \\ i \neq j}}^{K}L_{c_i,c_j} = 1, \quad \forall{c_i, c_j} \in {\bm{C}}
	\end{equation}
	\begin{equation}
	\label{out}
	\sum_{\substack{j=0 \\ j \neq i}}^{K}L_{c_i,c_j} = 1, \quad \forall{c_i, c_j} \in {\bm{C}}
	\end{equation}
	\begin{equation}
	\label{single}
	\sum_{c_i \in {\bm{F}}}\sum_{c_j \in \bm{F}}L_{c_i,c_j} \leq |\bm{F}| - 1,\quad \forall{\bm{F}} \subset \bm{C}; |\bm{F}| \geq 2.
	\end{equation}
	The constraints (\ref{in}) and (\ref{out}) guarantee that there is only one UAV path entering and leaving a given node,  which means that the UAV should visit each point in $\bm{C}$ exactly once. Constraint (\ref{single}) is the sub-trajectories elimination constraint and enforces that no partial loop exists where $\bm{F}$ is the subset of $\bm{C}$ \cite{R. Roberti}, which means there is only one single trajectory covering all CHs.
	
	According to the above analysis, the total energy consumption of the UAV is composed of the flying-related and the hovering-related energy consumption, which can be written as
	\begin{equation}
	\label{uav}
	    E_{\text{UAV}} = E_{\text{flight}} + \sum_{k=1}^{K}E_{c_k}.
	\end{equation}

	\subsection{Ground Network and Energy Model}
	We assume that all nodes have the same computation and transmission capabilities. In other words, all nodes  are capable of acting as a CH. Nodes are static after being deployed. All member nodes transmit their sensing information to CHs periodically and CHs forward the collected data to the UAV. We also assume that the transmission energy of each node is sufficient to send messages to its CH. In addition, the UAV can simultaneously connect to at most one CH. Hence, there is no interference among neighboring CHs.

	The energy consumption in the ground network includes two components. The first component is the communication energy consumption between CHs and their member nodes. The first-order radio model \cite{W. R. Heinzelman} is used to calculate the energy consumption of the ground network. The transmission energy is consumed by the transmitter's circuitry and power amplifier. If the distance between a member node and its CH is less than a given threshold, the power amplifier uses the free space model; otherwise, the multi-path model is used \cite{T. Rappaport}. The energy consumed to transmit an $l$-bit message from a member node to its CH $b_k$ is given by \cite{W. R. Heinzelman}
	\begin{equation}
	\label{groundenergy}
	E_{n}^{b_k} = lE_{\text{elec}} + l\left(\chi \varepsilon_\text{fs}d_{n,b_k}^2 + \left(1 - \chi \right) \varepsilon_\text{mp}d_{n,b_k}^4 \right)
	\end{equation}
	where
	\begin{equation}
	\chi = \left\{ \begin{array}{ll}
	1, & d_{n,b_k} \leq d_0 \\
	0, & d_{n,b_k} > d_0
	\end{array} \right.
	\end{equation}
	and
	\begin{equation}
	d_0 = \sqrt{\frac{\varepsilon_\text{fs}}{\varepsilon_\text{mp}}}.
	\end{equation}
	In (\ref{groundenergy}), $E_{\text{elec}}$ is the dissipated energy per bit in the circuitry, $d_{n,b_k}$ is the distance between the CH $b_k$ and one of its member nodes $n$, $n=1,\dots,N-1$, $d_0$ is the distance threshold, $\varepsilon_\text{fs}$ and $\varepsilon_\text{mp}$ represent the radio amplifier's energy parameter of the free space and multi-path fading models, respectively. Moreover, the energy consumed to receive an $l$-bit message from member node $n$ by $b_k$ is given by \cite{W. R. Heinzelman}
	\begin{equation}
	E_{b_k}^{(n)} = lE_{\text{elec}}.
	\end{equation}
	In addition, the second component of the energy consumption of the ground network is the energy consumed by each CH $b_k$ to complete its data transmission to the UAV. This component can be written as
	\begin{eqnarray}
	   E_{b_k} = P_{\text{CH}}T_k = P_{\text{CH}}\frac{(N-1)l}{r_{\text{data}}}
	\end{eqnarray}
	where $(N-1)l$ is the amount of data transferred by $b_k$ to the UAV. Hence, the total energy consumption of all nodes in the ground network in a complete data collection task, where member nodes transmit the sensing data to their CHs and CHs forward data to the UAV, is
	
	\begin{equation}
	\label{ground}
	E_\text{ground}  =  \sum_{k=1}^{K}\sum_{\substack{n=1}}^{N-1}\left(E^{b_k}_n + E_{b_k}^{(n)}\right) + \sum_{k=1}^{K}E_{b_k}.
	\end{equation}

	\subsection{Problem Formulation for UAV's Trajectory}
	Based on ($\ref{uav}$) and (\ref{ground}), the total weighted energy consumption in the UAV-WSN system can be formulated as
	\begin{align}
	\label{eq19}
	    E
		&=\omega \left(\sum_{k=1}^{K}\sum_{\substack{n=1}}^{N-1}\left(E^{b_k}_n + E_{b_k}^{(n)}\right) + \sum_{k=1}^{K}E_{b_k}\right) \nonumber \\
		&+ (1-\omega) \left(E_{\text{flight}} + \sum_{k=1}^{K}E_{c_k}\right), \,0 \leq \omega \leq 1
	\end{align}
	where the first term corresponds to the total energy consumption of the ground network, while the second term is the energy consumption of the UAV, and $\omega$ is the weighting coefficient that can be adjusted to achieve the trade-off between the two terms. With the aim of minimizing the total energy consumption of the ground network and the UAV, we jointly find a set of CHs from the ground cluster-based WSN and design the UAV's visiting order to these CHs. The optimization problem of interest is formulated as
	\begin{eqnarray}
	\label{objectivefunction}
	&&\min_{\substack{\{b_0,b_1,\dots,b_k,\dots, b_K\}\\ b_k\in \bm{G}_{k}}} \quad  E \nonumber\\
	\\
	&&\text{s.t.} \, (\ref{visit}), (\ref{in})-(\ref{single}).\nonumber
	\end{eqnarray}
	
	Obviously, the problem at hand is a constrained combinatorial optimization problem, which is NP-hard. Some promising approaches have been put forward to solve such combinatorial optimization problems, and their advantages and disadvantages are discussed below.
	\begin{enumerate}
		\item \emph{Exact methods}:  Exact methods often search for the optimal solution of the problem through systematic enumeration, integer programming, and constraint programming, etc.\cite{J. Puchinger and G. R. Raidl}. At least in theory, they can provide the optimal solution for the optimization problem. However, such algorithms cannot be applied to combinatorial optimization problems with large data scale because their computation complexity becomes prohibitive.
		
		\item \emph{Heuristics}: Heuristics are higher-level problem-independent algorithmic frameworks that provide a set of guidelines to develop optimization algorithms \cite{K. Sorensen}. However, they generally cannot guarantee to find globally optimal solutions.
		
		\item \emph{RL}: Q-learning, one of the RL techniques, is demonstrated to be promising in solving NP-hard problems \cite{X. Liu}. Specifically, it can deal with the path decision problem when provided with sufficient state space variables. However, if the number of ground nodes is high, Q-learning will need more storage space for action and state space variables \cite{B. Zhang}.

	\end{enumerate}
	
	
	As discussed before, DRL has recently shown to have important advantages in solving combinatorial optimization problems. A typical neural combinatorial framework is proposed in \cite{I. Bello} that uses RL to optimize a policy modeled by the pointer network. In \cite{M. Nazari}, the authors view a combinatorial optimization problem as a sequence of decisions, and they use a sequence-to-sequence neural network model and the RL approach to obtain a near-optimal solution for the optimization problem. Inspired by these promising developments, we extend the application of sequence-to-sequence model to solve the UAV's trajectory planning problem described earlier.
	
	\section{Deep Reinforcement Learning for UAV Trajectory Planning}\label{SecIV}
	
	Because all clusters $\{\bm{G}_1,\dots, \bm{G}_K\}$ must be visited by the UAV sequentially, we convert the visiting decisions problem into a sequence-to-sequence prediction problem. The problem can be simply formalized as follows. Given start position and all clusters, denoted by $\bm{\mathcal{G}} = \{b_0, \bm{G}_1,\dots, \bm{G}_K\}$, we want to output a permutation of the items in $\bm{\mathcal{G}}$ that maximizes some measure of interest. The output sequence is denoted as $\bm{\mathcal{T}} = \{\pi_0,\pi_1,\dots,\pi_K\}$, where each $\pi_t$ is the index of any element in $\bm{\mathcal{G}}$ being placed at the $t$-th position of $\bm{\mathcal{T}}$. In fact, $\bm{\mathcal{T}}$ is the UAV's visiting order to clusters in our problem. Thus, for a given input sequence $\bm{\mathcal{G}}$, the probability of the output sequence $\bm{\mathcal{T}}$ can be factorized by a product of conditional probabilities according to the chain rule
    \begin{equation}
    \begin{aligned}
    \label{chainrule}
        P_\theta(\bm{\mathcal{T}}|\bm{\mathcal{G}}) & = \prod_{t=0}^{K}P(\pi_t|\pi_0,\dots,\pi_{t-1}, \bm{\mathcal{G}})
    \end{aligned}
    \end{equation}
    where $t$ is the time step, $P_{\theta}(\bm{\mathcal{T}}|\bm{\mathcal{G}})$ parameterized by $\theta$ is a \emph{stochastic policy} for deciding the visiting order. The conditional probability $P(\pi_t|\cdot)$ models the probability of any cluster being visited at the $t$-th time step according to the given $\bm{\mathcal{G}}$ and clusters already visited at previous time steps \cite{I. Bello etc.}. A trained $\theta$ can assign high probabilities to good results and low probabilities to bad results. The reinforcement learning can be applied to train the optimal model policy $\theta^*$ for producing the optimal visiting order $\bm{\mathcal{T}}^*$ with the highest probability.

    \subsection{Pointer Network-A* Architecture for UAV's Trajectory Planning}

	

     \begin{figure*}[t]
		\centering
		\includegraphics[width=0.99\linewidth]{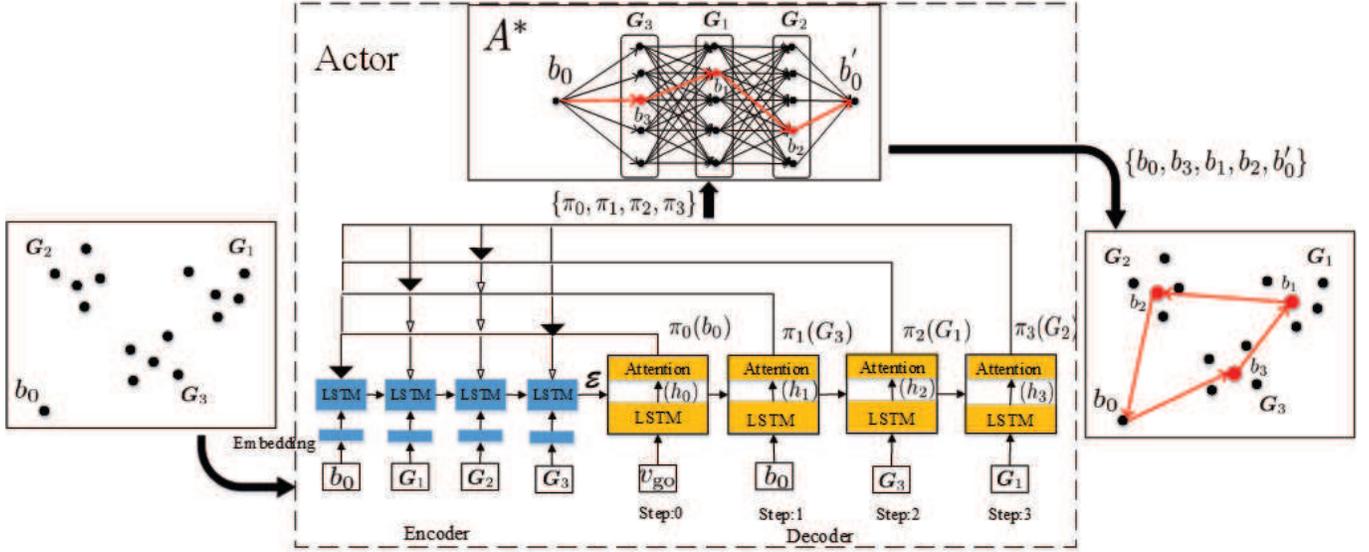}
		\caption{Example of Ptr-A* architecture for a 3-clusters network.}
		\label{figure2}
	\end{figure*}

    With the rapid development of neural network techniques, the neural network-based frameworks have been applied to sequence-to-sequence learning \cite{H. Zhang}. The general sequence-to-sequence neural network\cite{M. Nazari} encodes the input sequence into a vector that includes information of the input by a recurrent neural network (RNN), called encoder, and decodes the vector to the target sequence by another RNN, called decoder \cite{O. Vinyals}. In this work, we employ the pointer network to model the conditional probability $P(\pi_t|\cdot)$, which has been proved to be effective to solve the combinatorial optimization problems. The architecture of the pointer network is similar to sequence-to-sequence network model, but it uses attention mechanism as a pointer to choose items of its input sequence as the output.
    The proposed Ptr-A* model in this work is elaborated as follows.

    \subsubsection{Encoder}
    The  traditional RNN shows poor performance in dealing with the problem of long-term dependencies, which makes it difficult to be trained in practice\cite{Y. Bengio}. Hence, we use Long Short-Term Memory (LSTM) cells which are capable of learning long-term dependencies to construct a RNN as the encoder. Each item (a start position or a cluster) in $\bm{\mathcal{G}}$ is converted into a high $D$-dimensional vector space, which enables the policy to extract useful features much more efficiently in the transformed space\cite{W. Koehrsen}. Then, the embedding vectors are fed into LSTM cells. At each encoding step, the LSTM cell reads one embedded item and outputs a latent memory state. Finally, the input sequence $\bm{\mathcal{G}}$ is transformed into a sequence of latent memory states $\bm{\mathcal{E}} = \{e_0, \dots, e_K\}$, each $e_k\in\mathbb{R}^D$. In fact, the motive of the encoder network is to acquire the representation for each element in $\bm{\mathcal{G}}$.

    \subsubsection{Decoder}
    We also adopt LSTM cells to construct the RNN of the decoder network. The output $\{e_0, \dots, e_K\}$ of the encoder are given to the decoder network. At each decoding step $t$, the LSTM cell outputs the hidden state $h_t\in\mathbb{R}^D$ that includes the knowledge of previous steps. And then, the decoder employs the attention mechanism to output the visiting decision $\pi_t$ based on $h_t$ and $\{e_0, \dots, e_t\}$. Attention mechanism can help the model to give different weights to different elements of the input and extract more critical information \cite{O. Vinyals A. Toshev}. It tells us the relationship between each element in the input at current step $t$ and the output $\pi_{t-1}$ of the last decoding step. The most relevant element with the maximum conditional probability is selected as the access element at decoding step $t$. Thus, the calculation is given by
    \begin{eqnarray}
            \label{mu}
            u^t_j = \left\{ \begin{array}{ll}
	        \varphi \tanh{\left(W_{1}e_j + W_2h_t\right)}, & \text{if}\, j \notin \{\pi_0,\dots,\pi_{t-1}\} \\
	        -\infty, & \text{otherwise}
	\end{array} \right.
    \end{eqnarray}
    where $W_1$, $W_2\in\mathbb{R}^{D\times D}$ are attention matrices, $\varphi\in\mathbb{R}^{1\times D} $ is the attention vector. $W_1$, $W_2$, and $\varphi$ are denoted collectively by $\theta$, which is the learnable parameter in our pointer network. In essence, $u^t_j$ is the score associated with item $j$ ($e_j$) in position $t$.

    The conditional probability is calculated by a \emph{softmax} function over the remaining items (not visited in the previous steps), as follows:
    \begin{align}
    \label{pro}
        P(\pi_{t}&=j|\pi_0,\dots,\pi_{t-1}, \bm{\mathcal{G}})  = \text{softmax}\left(u^t_j\right) \nonumber \\
        & = \frac{\exp{\left(u^t_j\right)}}{\sum_{m\notin \{\pi_0,\dots,\pi_{t-1}\}}\exp{\left(u^t_m\right)}},\, j\in m.
    \end{align}
    The probability $P\left(\pi_{t}=j|\cdot\right)$ represents the degree to which the model points to item $j$ at the decoding step $t$ \cite{I. Bello etc.}. As shown in the example in Fig. \ref{figure2}, the start position $b_0$ and clusters $\bm{G}_1, \bm{G}_2, \bm{G}_3$ are inputted into the encoder network. $v_{\text{go}}$ is the start tag of the decoder, which is a learned vector. At each decoding step, the item of the input sequence with the highest probability is pointed by a thicker black arrow. The output of the 0-th decoding step points to $b_{0}\, (\pi_0)$, which will be visited at this step and given as the input of the next decoding step. Finally, we will obtain a visiting order sequence $\bm{\mathcal{T}} = \{\pi_0,\pi_1,\pi_2,\pi_3\}$ corresponding to the input sequence. 
    
    To help the reader familiarize with the attention mechanism, a numerical example is provided in Appendix A.

    \subsubsection{A* search}
    Once the output sequence of the decoder network is obtained, we can build a search graph for all clusters according to this sequence, where each layer is composed of nodes of one cluster. This is illustrated with an example in Fig.~ \ref{figure2}. It is worth mentioning that the first layer is the start position $b_0$ and the last layer is the end position $b_0^{'}$ which is the copy of $b_0$. Thus, the created graph has a total of (K+2) layers. We use the A* search algorithm, one of best path-finding algorithms, to find the CH from each cluster to build a path having the smallest cost (total weighted energy consumption of the UAV-WSN system) from the start position to the end position. In each iteration, the A* algorithm needs to calculate the cost of the traversed path and the estimated cost required to extend the path to the end to determine which of its partial paths to expand into one or more longer paths\cite{V. Razo and H. Jacobsen}. Any node $m$ is chosen to be visited by the following function
    \begin{equation}
        f(m) = g(m) + h(m)
    \end{equation}
    where $g(m)$ represents the exact energy consumption of the UAV-WSN system when the UAV moves from the start node to a candidate node $m$, following the path generated to get there, $h(m)$ is the estimated energy consumption of the UAV to travel from the candidate node $m$ to the end. Then, the node with the lowest $f(m)$ value is selected from candidate nodes as the next node to be traversed.

    The main implementation of the A* algorithm is to maintain two lists. The OPEN list contains those nodes that are candidates for checking.  The CLOSED list contains those nodes that have been checked. The neighbor nodes of any node located in any layer are defined as all nodes in its previous and next layers. Also, each node keeps a pointer to its parent node so that we can determine how it was found, which is implemented by a map COME\_FROM. The pseudocode of using the A* algorithm to find the path from the start position to the end position is described in Algorithm 1. As shown in the example in Fig.~\ref{figure2}, the output of the A* algorithm is the trajectory from $b_0$ to $b_0^{'}$, which can ensure the minimum energy consumption $E$. In addition, the CH of each cluster is found on this trajectory, given by $\{b_0, b_3, b_1, b_2, b_0^{'}\}$. Finally, the Ptr-A* model outputs the trajectory and the minimum energy consumption $E$.

    \begin{algorithm}[t!]
	\caption{A* search algorithm for the trajectory planning}
	\label{alg1}
	\begin{algorithmic}[1]
		\renewcommand{\algorithmicrequire}{\textbf{Input:}}
		\renewcommand{\algorithmicensure}{\textbf{Output:}}
		\REQUIRE $\bm{\mathcal{T}}$
	    \ENSURE  Trajectory, minimum energy consumption $E$\\
	    \STATE Build a search graph by $\bm{\mathcal{T}}$
	    \STATE Initialize OPEN, CLOSED, and COME\_FROM\\
	    \STATE $f(b_0) = 0$, OPEN.add$(b_0)$\\
	     \WHILE{OPEN is not empty}
	         \STATE Find the node $q$ with the lowest $f(q)$ from OPEN\\
	         \IF{$q=b_0^{'}$}
	         \STATE Construct path from $b_0$ to $b_0^{'}$ by COME\_FROM\\
	         \RETURN Trajectory, $E$
	         \ENDIF
	         \STATE OPEN.remove($q$)\\
	         \STATE CLOSED.add($q$)\\
	         \STATE Obtain neighbor nodes of $q$ \\
	         \FOR{each neighbor node $m$ of $q$}
	             \STATE $cost = g(q)$ + the energy consumption of UAV-WSN from $q$ to $m$\\
	             \IF{$m$ in OPEN and $cost < g(m)$ }
	                \STATE OPEN.remove($m$)
	             \ENDIF
	             \IF{$m$ in CLOSED and $cost < g(m)$ }
	                \STATE CLOSED.remove($m$)
	             \ENDIF
	              \IF{$m$ not in OPEN and CLOSED}
	                 \STATE $g(m) = cost$\\
	                 \STATE $f(m) = g(m) + h(m)$\\
	                 \STATE OPEN.add($m$)
	                  \STATE COME\_FROM[$m$]= $q$ //set $m$'s parent
	              \ENDIF
	         \ENDFOR
	     \ENDWHILE
	\end{algorithmic}
    \end{algorithm}

	\subsection{Parameters Optimization with Reinforcement Learning}
	 In order to find a good trajectory for the UAV, we need to obtain the optimal model parameter $\theta^*$ that can be trained from samples. If we adopt a supervised learning to train the model parameter, high-quality labeled data is needed because it decides the performance of the model. However, it is expensive to get the high-quality labeled data in practice for the proposed UAV's trajectory problem. Instead, we choose the well-known model-free policy-based RL, known as the actor-critic algorithm \cite{V. R. Konda}, to train the model because it is shown to be an appropriate paradigm for training neural networks for combinatorial optimization \cite{I. Bello}. The UAV works as the agent to make a sequential action set in a given state of the environment. In the following, we describe the state, action, reward, and training of the proposed DRL algorithm.
	 \subsubsection{State}
	 The state includes coordinates for all clusters,  the UAV's location, and the energy consumption of UAV-WSN at current step $t$.
	 \subsubsection{Action} The action represents the choice of the next cluster to be selected at current step $t$ and the CH in this cluster. Thus, we define the output of the right-hand side of (\ref{chainrule}) and the CH selection by A* as the action at each step.
	 \subsubsection{Reward} We design the reward as the negative of the total energy consumption in (\ref{eq19}). This means that the DRL is set to get the maximal reward (minimal energy consumption).

	 \subsubsection{Training}
	
	 The actor-critic method includes the actor network and the critic network. The actor network is the proposed Ptr-A* in this work. The critic network is used to provide an approximated baseline of the reward for any problem instance to reduce the variance of gradients during the training phase and increases the speed of learning \cite{V. R. Konda}. Our critic network, parameterized by $\psi$ , has the same architecture as that of the encoder of the Ptr-A*. Then, its hidden states are decoded into a baseline prediction by two fully-connected ReLU layers \cite{I. Bello}.
	\begin{algorithm}[t!]
	\caption{Training Ptr-A* by Actor-Critic algorithm}
	\label{alg1}
	\begin{algorithmic}[1]
		\renewcommand{\algorithmicrequire}{\textbf{Input:}}
	     \REQUIRE Training samples set $\bm{D}=\{\bm{\mathcal{G}}_1, \bm{\mathcal{G}}_2, \dots\}$, batch size $B$, training steps $S$
	    \STATE Initialize actor network $\theta$ and critic network $\psi$ with random weights\\
	    \FOR{$s$ = 1 to $S$}
		\STATE Sample $\bm{\mathcal{G}}_i$ from $\bm{D}$, $\forall{i}\in\{1,\dots, B\}$
		\STATE Calculate $E_i$ and $\bm{\mathcal{T}}_i$ with Ptr-A* network, $\forall{i}\in\{1,\dots, B\}$
		\STATE Calculate  $V_{\psi}\left(\bm{\mathcal{G}}_i\right)$ with Critic network, $\forall{i}\in\{1,\dots, B\}$
		\STATE $d\theta \gets \frac{1}{B} \sum_{i=1}^B \left( E_i-V_{\psi}\left(\bm{\mathcal{G}}_i\right) \right) \nabla_{\theta} \log p_{\theta} \left(\bm{\mathcal{T}}_i|\bm{\mathcal{G}}_i\right)$
		\STATE $L(\psi) \gets \frac{1}{B}\sum_{i=1}^{B}\left(V_{\psi}\left(\bm{\mathcal{G}}_i\right)-E_i \right)^2$
		\STATE $\theta \gets \text{Adam}\left(\theta, d\theta\right)$
		\STATE $\psi \gets \text{Adam}\left(\psi, \nabla_\psi L(\psi)\right)$
		\ENDFOR
		\RETURN $\theta^* = \theta$
	\end{algorithmic}
    \end{algorithm}
    Our training objective is the expected energy consumption, which is defined as
	\begin{equation}
	    \label{J}
	    J\left(\theta|\bm{\mathcal{G}}\right) = \mathbb{E}_{\bm{\mathcal{T}}\sim p_{\theta(.|\bm{\mathcal{G}})}} [E].
	\end{equation}
	We use policy gradient method and stochastic gradient descent to optimize $\theta$. The gradient of (\ref{J}) is formulated by REINFORCE\cite{R. J. Williams} algorithm
	\begin{equation}
	\label{gradient}
	    \nabla_{\theta}J\left(\theta|\bm{\mathcal{G}}\right) = \mathbb{E}_{\bm{\mathcal{T}}\sim p_{\theta(.|\bm{\mathcal{G}})}}\left[\left(E - V_{\psi}\left(\bm{\mathcal{G}}\right)\right) \nabla_{\theta}\log p_{\theta}\left( \bm{\mathcal{T}} | \bm{\mathcal{G}}\right) \right]
	\end{equation}
	where $V_{\psi}\left(\bm{\mathcal{G}}\right)$ is a baseline function for reducing the variance of the gradients, which is implemented by the critic network. Assume we have $B$ $i.i.d$ train samples, the gradient in (\ref{gradient}) can be approximated with Monte Carlo sampling as follows
	\begin{equation}
	    \nabla_{\theta}J\left(\theta\right) \approx \frac{1}{B} \sum_{i=1}^B \left( E_i-V_{\psi}\left(\bm{\mathcal{G}}_i\right) \right) \nabla_{\theta} \log p_{\theta} \left(\bm{\mathcal{T}}_i|\bm{\mathcal{G}}_i\right).
	\end{equation}
	 We train the parameters of the critic with stochastic gradient descent on a mean squared error objective $L(\psi)$ between its predictions $V_\psi \left(\bm{\mathcal{G}}_i \right)$ and the actual energy consumption. $L(\psi)$ is formulated as
	 \begin{equation}
	     L(\psi) = \frac{1}{B} \sum_{i=1}^B\left(V_{\psi}\left(\bm{\mathcal{G}}_i\right)-E_i\right)^2.
	 \end{equation}
	 The training procedure is presented in Algorithm \ref{alg1}. Notice that Adam algorithm is used to update the parameters of the actor network and the critic network iteratively. Adam algorithm designs independent adaptive learning rates for different parameters via calculating the first and second moment estimates of the gradient instead of using a single learning rate to update all parameters by the traditional random gradient descent \cite{D. P. Kingma}. Given the initial learning rate, the learning rates in different steps adaptively change according to the learning results. Because of the generalization property of RNNs \cite{Z. Tu}, our proposed models, including the Ptr-A* and critic, have a very good generalization ability. In the training phase, we can use small problem instances to train the model, and then the trained model can be utilized to solve large problem instances.
	
\section{Numerical Results}\label{SecVI}

\begin{table}[t!]
	\centering
	\caption{Simulation parameters}
	\label{table1}
	\scriptsize
	 \begin{tabular}{p{1.5cm}<{\centering}|p{3.9cm}<{\centering}|p{2.3cm}<{\centering}}
		\hline{}
		\textbf{Parameter} & \textbf{Description} & \textbf{Value}\\
		\hline
		$\varepsilon_\text{fs}$ & Amplifier’s  energy  parameter of the free space fading & 10 $\text{pJ/bit/m}^2$ \cite{W. R. Heinzelman}\\
		\hline
		$\varepsilon_\text{mp}$ & Amplifier’s  energy  parameter of the multi-path fading & 0.0013 $\text{pJ/bit/m}^2$ \cite{W. R. Heinzelman}\\
		\hline
		$E_{\text{elec}}$ & Energy consumption per bit in the circuitry & 50 $\text{nJ/bit}$ \cite{W. R. Heinzelman}\\
		\hline
		$P_{\text{CH}}$ & Transmit power of each CH & 21 $\text{dBm/Hz}$ \cite{M. B. Ghorbel}\\
		\hline
        $N$ & Number of nodes per cluster & 20\\
        \hline
        $B_{\text{width}}$ & Bandwidth & 1 MHz \\
        \hline
        $N_0$ & Noise power & $-174$ dBm/Hz \cite{M. B. Ghorbel}\\
        \hline
        $f_c$ & Carrier frequency & 2 GHz \cite{M. B. Ghorbel}\\
        \hline
        $\alpha$ & Path loss exponent & 3 \cite{M. B. Ghorbel}\\
        \hline
        $H$ & UAV's flight height & 50 m \\
        \hline
        $\mu_{\text{LoS}}$, $\mu_{\text{NLoS}}$ & Mean values of the excessive path loss & 1 $\text{dB}$, 20 $\text{dB}$ \cite{A. Al-Hourani}\\
        \hline
        $\beta, \eta$ & Environmental parameters &  0.03, 10 \cite{M. B. Ghorbel}\\
        \hline
        $v_{\text{UAV}} = v_{\text{max}}$ & UAV's flight speed & 15 $\text{m/s}$ \cite{M. B. Ghorbel}\\
        \hline
        $m_{\text{tot}}$ & UAV's mass & 500 g \cite{H. Ghazzai}\\
        \hline
        $r_p$ & Radius of UAV's propellers & 20 cm \cite{H. Ghazzai}\\
        \hline
        $n_p$ & Number of propellers & 4 \cite{H. Ghazzai}\\
        \hline
        $P_{\text{max}}$ &  UAV's hardware power level at full speed & 5 W \cite{H. Ghazzai}\\
        \hline
        $P_{\text{idle}}$ & UAV's hardware power level when it hovers & 0 W \cite{H. Ghazzai}\\
        \hline
        $P_{\text{com}}$ & UAV's communication power & 0.0126 W \cite{M. B. Ghorbel}\\
        \hline
		\end{tabular}
\end{table}

In this section, we first introduce detailed environment settings, and then describe the decoding search strategies at inference. Furthermore, we compare the performance of the proposed DRL algorithm with several baseline algorithms.

\subsection{Environmental Settings and Model Training}

We consider the ground network size of 2 km $\times$ 2 km, and the start position of the UAV is located at $(0\,\text{m}, 0\,\text{m})$. Simulation parameters are listed in Table \ref{table1}. We use mini-batches of size 512 and LSTM cells with 128 hidden units in the encoder and the decoder. We implement the proposed model by using Pytorch 1.4 and Python 3.7 on a VM instance of Google Cloud Platform with 1 NVIDIA TESLA P100 GPU. The parameters of both the actor and critic networks are initialized by the Xavier initialization method and trained by the Adam optimizer with an initial learning rate of 0.0001 and decayed every 5,000 steps by 0.96.

{It is assumed that nodes in a given cluster $\bm{G}_k$ are distributed according to the Gaussian distribution. Each cluster's nodes are} sampled from a \emph{torch.normal}$(\nu, \text{std})$ where $\nu$ is the mean and std is the constant standard deviation. Each Gaussian distribution's $\nu$ is randomly sampled from a \emph{torch.rand()} function to determine the position of each cluster in the two-dimensional space. We train the model using instances of 20 clusters and 40 clusters, respectively. The 20-clusters model is trained for 100,000 steps, and the 40-clusters model is trained for 200,000 steps. We give a simple example on how to obtain the train data. At each training step of the 20-clusters model, we sample 20 means from \emph{torch.rand()} for 20 Gaussian distributions, respectively. Then, we use these 20 \emph{i.i.d.} distributions to generate a set (problem instance) of 20 clusters where the number of nodes per cluster is 20. The test data sets are also generated in the same way, only the number of clusters is different.

\subsection{Decoding Search Strategies at Inference}
Given a new problem instance $\bm{\mathcal{G}}$ at inference, the decoder network of our trained Ptr-A* architecture can easily output an access sequence for all clusters. The decoding process of the decoder at inference shows how solvers search over a large set of feasible access sequences. In this work, we consider the following three decoding search strategies.

\subsubsection{Greedy Search}
Greedy search strategy always select the cluster with the largest probability at each decoding step during inference, which is labeled as DRL-greedy in the simulation results.

\subsubsection{Sampling Search}
This strategy samples $M$ candidate solutions from the \emph{stochastic policy} $ P_\theta(\cdot|\bm{\mathcal{G}})$ by running the trained Ptr-A* on a single test input $\bm{\mathcal{G}}$ and selects the one with the minimum expected energy consumption from $M$ candidate outputs. The more we sample, the more likely we will get the better output. In the simulation, we set $M = 51200$, and this strategy is labeled as DRL-sampling.

\begin{algorithm}[t!]
	\caption{Active Search}
	\label{active}
	\begin{algorithmic}[1]
		\renewcommand{\algorithmicrequire}{\textbf{Input:}}
	     \REQUIRE Test input $\bm{\mathcal{G}}$, steps $S$, $\zeta$
	    \STATE Randomly sample a solution $\bm{\mathcal{T}}$ for $\bm{\mathcal{G}}$\\
	    \STATE Calculate $E$ according to $\bm{\mathcal{T}}$ by A*
	    \STATE $O \gets E$
	    \FOR{$s$ = 1 to $S$}
		\STATE  $\bm{\mathcal{T}}_i$ $\sim$ Sample solutions $P_{\theta}(\cdot|\bm{\mathcal{G}})$, $\forall{i}\in\{1,\dots, Q\}$
		\STATE  $E_{(\bm{\mathcal{T}}_j|\bm{\mathcal{G}})}\gets$ argmin$\left(E_{(\bm{\mathcal{T}}_1|\bm{\mathcal{G}})}, \dots, E_{(\bm{\mathcal{T}}_Q|\bm{\mathcal{G}})} \right)$
		\IF{$E_{(\bm{\mathcal{T}}_j|\bm{\mathcal{G}})} < E$}
		  \STATE $\bm{\mathcal{T}} \gets \bm{\mathcal{T}}_j$\\
		  \STATE $E \gets E_{(\bm{\mathcal{T}}_j|\bm{\mathcal{G}})}$\\
		\ENDIF
		\STATE $d\theta \gets \frac{1}{Q} \sum_{i=1}^Q \left( E_{(\bm{\mathcal{T}}_i|\bm{\mathcal{G}})}-O \right) \nabla_{\theta} \log p_{\theta} \left(\bm{\mathcal{T}}_i|\bm{\mathcal{G}}\right)$
		\STATE $\theta \gets \text{Adam}\left(\theta, d\theta\right)$
		\STATE $O \gets \zeta O + (1-\zeta) \frac{1}{Q}\sum_{i=1}^Q V_{\psi}\left(\bm{\mathcal{G}}\right)$
		\ENDFOR
		\RETURN $\bm{\mathcal{T}}, E$
	\end{algorithmic}
\end{algorithm}

\subsubsection{Active Search}
Unlike the greedy search and the sampling search, this strategy can refine the parameter $\theta$ of the Ptr-A* during inference to minimize the expected energy consumption on a single test input $\bm{\mathcal{G}}$. Active search samples multiple solutions $\bm{\mathcal{T}}_1, \dots, \bm{\mathcal{T}}_Q$ from $ P_\theta(\cdot|\bm{\mathcal{G}})$ for a single test input $\bm{\mathcal{G}}$ and uses policy gradients to refine $\theta$ \cite{I. Bello}. The process is presented in Algorithm \ref{active}. In the simulation, we sample three different sets of candidate solutions, $\{512, 5120, 10240\}$, which are labeled with DRL-active-512, DRL-active-5120, and DRL-active-10240, respectively.

\begin{figure}[!t]
  \centering
    \hspace{0in}\subfigure[]{\includegraphics[width=0.45\textwidth]{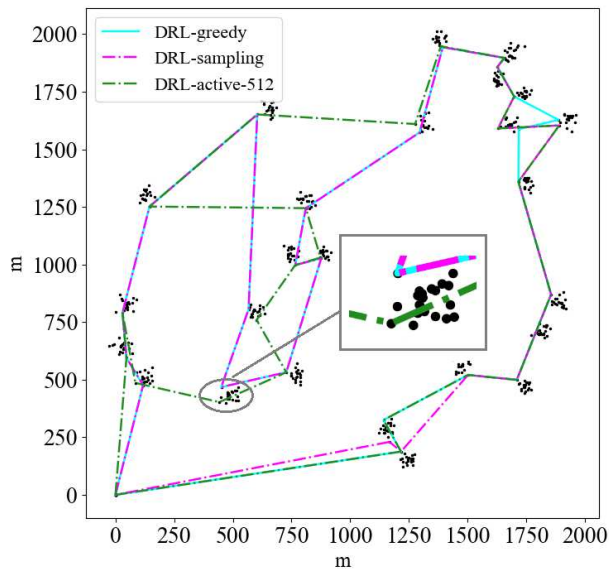}}\hspace{0in}
	 \subfigure[]{\includegraphics[width=0.45\textwidth]{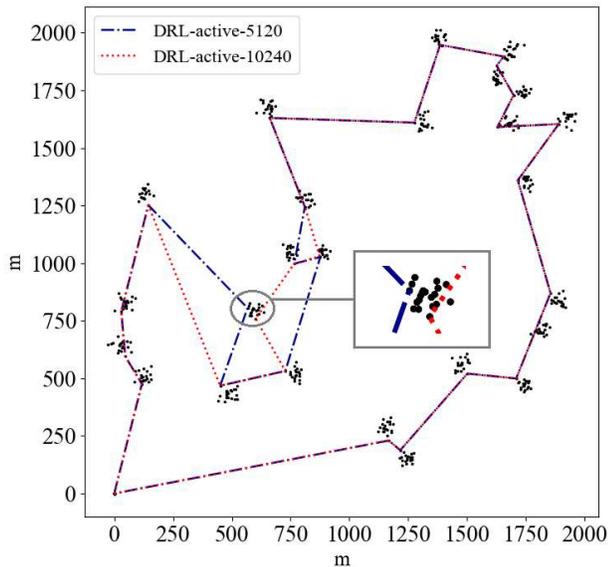}} \hspace{0in}
     \caption{Trajectories comparison on 25 clusters test instance when $\omega=0$.}
     \label{figure3}
	\vspace{0in}
\end{figure}

\subsection{Small-Scale Clusters}

To thoroughly  evaluate the performance of the proposed DRL algorithm, we first test the trained 20-clusters model on small-scale clusters. Since $\omega$ in (\ref{eq19}) is the weighting coefficient, its value does not impact the comparison results among algorithms. When $\omega = 0$, our optimization problem only considers the energy consumption of the UAV, which mainly depends on the flying distance of the UAV. Fig.~\ref{figure3} illustrates how the proposed DRL algorithm performs with different search strategies on the 25-clusters problem instance. As can be seen, the trajectory generated by DRL-greedy is the longest (9241 m), while DRL-active-10240 produces the shortest trajectory (8693 m) among strategies.

\begin{figure}[!t]
		\centering
		\includegraphics[width=1\linewidth]{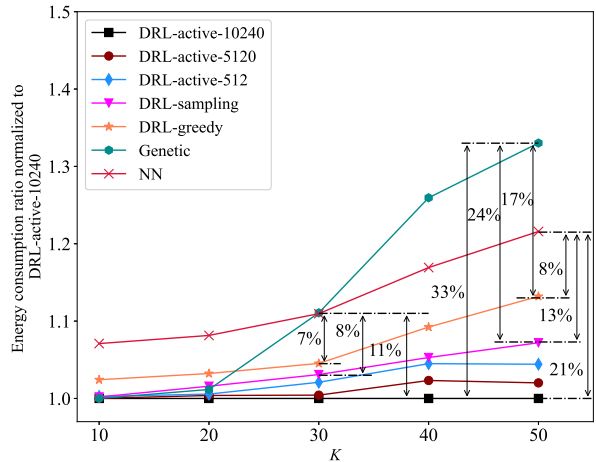}
		\caption{Energy consumption comparison on small-scale clusters.}
		\label{figure4}
\end{figure}

\begin{table}[t!]
	\centering
	\caption{Running time comparison on small-scale clusters.}
	\label{timesmall}
	 \begin{tabular}{p{2.75cm}<{\centering}|p{0.7cm}<{\centering}|p{0.7cm}<{\centering}|p{0.7cm}<{\centering}|p{0.7cm}<{\centering}|p{0.7cm}<{\centering}}
		\hline
	    & \multicolumn{5}{c}{Time (s)} \\
		\hline
	    \backslashbox{Algorithm}{$K$} & 10 & 20 & 30 & 40 & 50\\
		\hline
	    DRL-active-10240 & 17.1 & 57.15 & 121.62 & 209.27 & 319.64\\
		\hline
		DRL-active-5120 & 8.59 & 29.12 & 60.85 & 105.82 & 160.68\\
		\hline
		DRL-active-512 & 0.98 & 3.4 & 7.05 & 12.11 & 18.48\\
		\hline
		DRL-sampling & 14.26 & 29.34 & 43.39  & 59.89 & 81.3 \\
	    \hline
	    DRL-greedy & 0.31 & 0.63 & 1.07 & 1.68 & 2.46\\
	    \hline
	    Genetic & 60.37 & 64.85 & 73.35 & 80.21 & 90.91\\
	    \hline
	    NN & 1.12 & 1.12 & 1.13 & 1.15 & 1.16\\
	    \hline
		\end{tabular}
\end{table}

Next, we compare our proposed DRL algorithm having different decoding search strategies with the nearest neighbor (NN) heuristic \cite{B. Hu} and the genetic algorithm \cite{J. Li}. We first investigate the energy consumption comparison between our proposed DRL on the trained 20-clusters model and two baselines when $\omega=0.5$. The genetic algorithm runs for 4,000 generations, the chance of mutation is 0.5\%, and the size of population is 150. In Fig.~\ref{figure4}, we plot the average ratios of the energy consumption of our proposed DRL algorithm with different search strategies and two baselines to the energy consumption of DRL-active-10240 versus different numbers of clusters $K$. Although the model is trained on 20-clusters problem instances, it still obtains good performance on the 10-clusters, 30-clusters, 40-clusters, and 50-clusters networks. This shows that the proposed DRL algorithm achieves an excellent generalization ability with respect to the number of clusters used for training. When $K=10$, genetic, DRL-active-10240, DRL-active-5120, DRL-active-512, and DRL-sampling obtain almost the same energy consumption result; however, the NN algorithm has higher energy consumption when compared with our proposed DRL with active search and sampling search strategies. As the number of clusters increases, the energy consumption savings of our proposed algorithm increase when compared to the NN and genetic algorithms. For example, when $K=30$, the energy consumptions of UAV-WSN produced by NN and genetic algorithms are almost equal, which is about $11\%$ more than that of DRL-active-10240, $8\%$ more than that of DRL-sampling, and $7\%$ more than that of DRL-greedy. When the number of clusters increases to 50, the energy consumption of NN is around $21\%$ more than that of DRL-active-10240, $13\%$ more than that of DRL-sampling, and $8\%$ more than that of DRL-greedy. Likewise for $K=50$, the energy consumption of the genetic algorithm is around $33\%$ more than that of DRL-active-10240, $24\%$ more than that of DRL-sampling, and $17\%$ more than that of DRL-greedy. It can be seen that our proposed DRL algorithm using any of three active search strategies can achieve better results than other search strategies and algorithms. This is because the active search strategy can refine the parameters of the Ptr-A* model for producing the best solution while searching for candidate solutions on a single test instance at inference. From Fig.~\ref{figure4}, we can see that DRL-sampling also obtains a relatively competitive result.

\begin{figure}[!t]
  \centering
    \hspace{0.1in}\subfigure[On 40-clusters model.]{\includegraphics[width=0.485\textwidth]{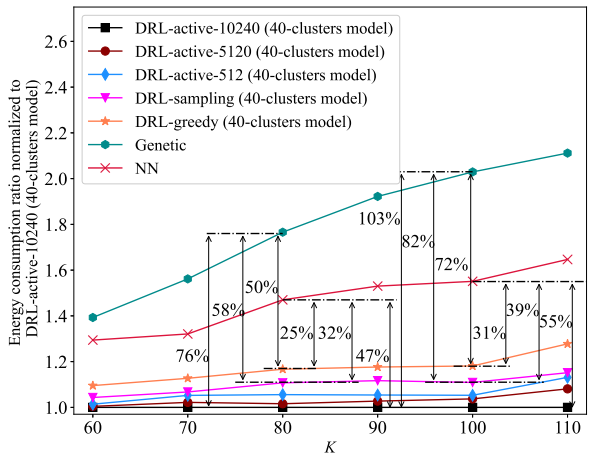}}\hspace{0in}
	 \subfigure[On 20-clusters model.]{\includegraphics[width=0.485\textwidth]{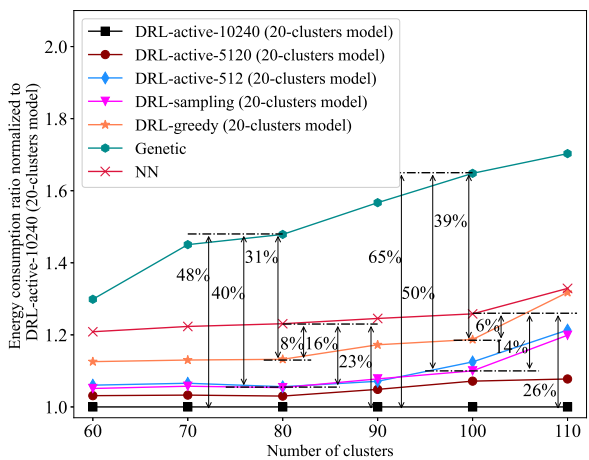}} \hspace{0in}
	 \subfigure[40-clusters model vs. 20-clusters model]{\includegraphics[width=0.485\textwidth]{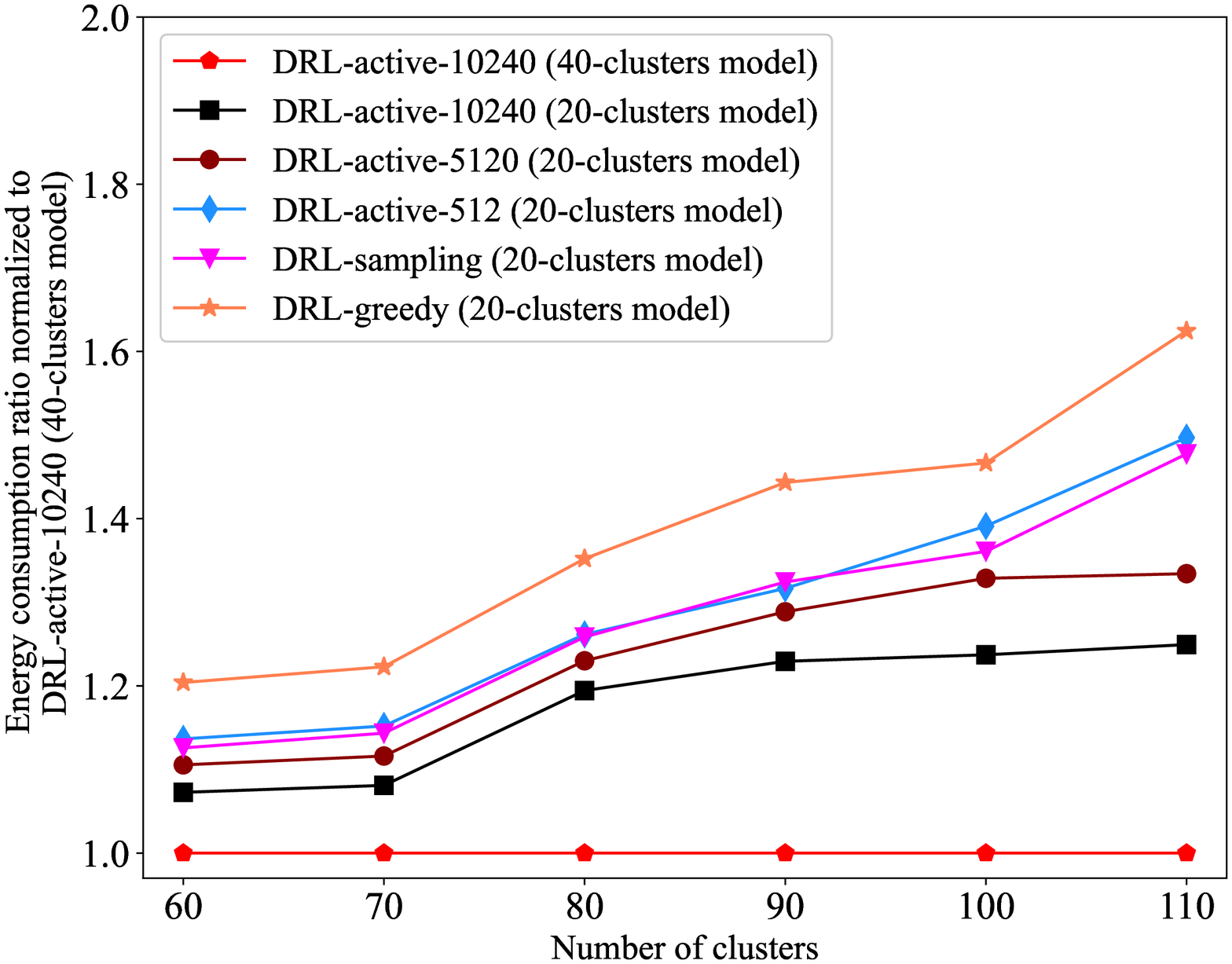}}
     \caption{Energy consumption comparison on large-scale clusters}
     \label{figure5}
\end{figure}

Table \ref{timesmall} compares the running time at inference. As the number of clusters increases, the running time of the proposed DRL algorithm with all strategies and two baseline techniques increases. Although DRL-active-10240 obtains the best performance in reducing the energy consumption as can be seen from Fig.~\ref{figure4}, it has the longest running time. This is because it needs more iterations to update the parameters. If the number of candidate solutions is relatively small, active search strategy tends to spend less time to produce the solution, like DRL-active-512. In addition, we can see that the running time of NN is always minimal among all algorithms for all values of $K$. DRL-greedy's running time is comparable with that of NN. Meanwhile, the running times of DRL-sampling and DRL-active-512 are lower when compared with the genetic algorithm.

\subsection{Large-Scale Clusters}

In this subsection, we test the performance of the trained models on large-scale clusters test instances. The genetic algorithm runs for 10,000 generations. We first observe the results of the proposed DRL algorithm on the trained 40-clusters model, as shown in Fig.~\ref{figure5} (a). Clearly, our proposed DRL algorithm exhibits much better performances than the two baseline techniques in reducing the energy consumption of the UAV-WSN system. As the number of clusters increases, there is an increasing performance gap between the proposed DRL algorithm and the baseline techniques. For instance, when $K=80$, the genetic algorithm consumes $50\%$, $58\%$, and $76\%$ more energy than when compared to DRL-greedy, DRL-sampling, and DRL-active-10240, respectively. As the number of clusters increases to 100, the energy consumption of the UAV-WSN when using the genetic algorithm is $72\%$ more than that of DRL-greedy, $82\%$ more than that of DRL-sampling, and $103\%$ more than that of DRL-active-10240. NN also shows a similar trend to that of the genetic algorithm. In particular, its energy consumption is $47\%$ more than that of DRL-active-10240 when $K=80$ and increases to $55\%$ more than the energy consumption of DRL-active-10240 when $K=100$. When compared with DRL-greedy, the extra amount of energy consumed by the UAV-WSN when using NN increases from $25\%$  to $31\%$ as the number of clusters increasing from 80 to 100. However, NN exhibits an obviously superior performance than the genetic algorithm.  As we can see, the active search strategies outperform the greedy strategy and sampling strategy for large-scale clusters problems. DRL-sampling shows a slightly better performance than DRL-greedy, which is reasonable.

Next, we use the trained 20-clusters model to evaluate the performance of the proposed DRL algorithm on large-scale clusters test instances. In Fig.~\ref{figure5} (b), our proposed DRL algorithm with different search strategies still obtains relatively good results as compared to the two baseline techniques on large-scale problem instances when the value of $K$ varies, while the results generated by active search strategies are still the best among all search strategies. For example, when $K=80$, the energy consumption of the genetic algorithm is $48\%$ more than that of DRL-active-10240, $40\%$ more than that of DRL-sampling, and $31\%$ more than that of DRL-greedy.  Although NN exhibits a better performance than the genetic algorithm, its energy consumption is $23\%$ more than that of DRL-active-10240, $16\%$ more than that of DRL-sampling, and $8\%$ more than that of DRL-greedy when $K=80$. However, compared with the results obtained on the trained 40-clusters model, the 20-clusters model clearly shows inferior performance. When $K=100$, the energy consumption gap between the genetic algorithm and DRL-active-10240 is $103\%$ on the 40-clusters model, but this gap decreases to $65\%$ on the 20-clusters model. Likewise, the gap in the energy consumption between the genetic algorithm and DRL-active-10240 decreases from $76\%$ on the 40-clusters model to $48\%$ on the 20-clusters model when $K=80$.

To investigate the difference between the two trained models, we compare the results obtained on the 20-clusters model with the results of DRL-active-10240 on the 40-clusters model. As shown in Fig.~\ref{figure5}~(c), DRL-active-10240 (40-clusters model) clearly exhibits superior performance in reducing the energy consumption than three search strategies on the 20-clusters model. This performance merit constantly increases as the value of $K$ increases. Thus, the trained 40-clusters model is more suitable for solving the trajectory planning problem for the UAV in cases of large-scale clusters.

\begin{table*}[!t]
	\centering
	\caption{Running time comparison on large-scale clusters.}
	\label{timebig}
	\begin{tabular}{p{2cm}<{\centering}| p{2.1cm} |p{0.7cm}<{\centering}|p{0.7cm}<{\centering}|p{0.7cm}<{\centering}|p{0.7cm}<{\centering}|p{0.8cm}<{\centering}|p{0.8cm}<{\centering}}
		\hline
	    \multicolumn{2}{c|}{} & \multicolumn{6}{c}{Time (s)} \\
		\hline
	    \multicolumn{2}{c|}{\backslashbox{Algorithm}{$K$}} & 60 & 70 & 80 & 90 & 100 & 110\\
		\hline
	    \multirow{2}*{DRL-active-10240} & 20-clusters model & 448.65 & 584.71 & 754.23 & 920.83 & 1149.36 & 1329.96\\
		\cline{2-8}
		 & 40-clusters model & 449.01  & 584.81 & 755.37 & 920.61 & 1150.05 & 1330.28\\
		\hline
		\multirow{2}*{DRL-active-5120} & 20-clusters model & 225.98  & 293.52 & 382.03 & 480.87 & 590.5 & 705.11\\
		\cline{2-8}
		 & 40-clusters model & 226.26 & 294.31 & 382.15  & 480.92 & 590.04 & 705.36\\
		\hline
		\multirow{2}*{DRL-active-512} & 20-clusters model & 24.99 & 33.08 & 44.08 & 55.54 & 67.21 & 79.31\\
		\cline{2-8}
		 & 40-clusters model & 25.75  & 33.57  & 44.1 & 55.78 & 67.54 & 79.88\\
		\hline
		\multirow{2}*{DRL-sampling} & 20-clusters model & 102.06 & 148.72 & 171.65 & 210.4 & 257.64 & 306.85\\
		\cline{2-8}
		 & 40-clusters model & 102.47  & 148.79 & 171.9 & 210.29 & 258.53 & 307.11\\
	    \hline
	    \multirow{2}*{DRL-greedy} & 20-clusters model & 3.11 & 4.23 &  5.84 & 7.83 & 9.96 & 12.55\\
	    \cline{2-8}
		 & 40-clusters model & 3.39  & 4.54 & 5.81 & 7.48 & 10.2 & 12.89\\
	    \hline
	    \multicolumn{2}{c|}{Genetic} & 213.39  & 217.06 & 225.5 & 233.88 & 239.54 & 247.19\\
	    \hline
	    \multicolumn{2}{c|}{NN} & 1.17 & 1.17 & 1.18 & 1.19 & 1.2 & 1.22\\
	    \hline
		\end{tabular}
\end{table*}

The running time comparison of different strategies and algorithms at inference on large-scale problem instances is provided in Table \ref{timebig}. We can observe that for a given number of clusters, the same strategy running on different models produces solution in almost the same amount of time. For example, DRL-active-10240 takes 448.65 s on the 20-clusters model and 449.01 s on the 40-clusters model. DRL-active-10240 takes the longest time among all the algorithms because it requires more iterations to refine the parameters of the Ptr-A* to produce the best performance. The running time of DRL-sampling is acceptable in comparison with the genetic algorithm given its performance merits. Furthermore, the computation time spent by DRL-greedy is the least among all strategies, which is also significantly less than the running time of the genetic algorithm and slightly more than the time spent by NN. Although the NN algorithm spends the least amount of time, it produces worse results.

\section{Conclusions}\label{SecVII}

In this paper, we investigated the problem of designing the UAV's trajectory for a clustered WSN to minimize the total energy consumption in the UAV-WSN system.
Inspired by the recent developments of DRL, we propose a novel DRL-based method to solve the UAV's trajectory planning problem. Because the visiting order of the UAV to clusters can be regarded as a sequential decision problem, we design a Ptr-A* model to produce the trajectory of the UAV. The pointer network of the proposed Ptr-A* model is used to determine the visiting order to clusters. Then, a search graph for all clusters is built according to the visiting order. The A* search algorithm is utilized to quickly find the CH for each cluster from the search graph with the aim of minimizing the energy consumption of the UAV-WSN system. In order to obtain optimal parameters of the Ptr-A*, we employ the model-free RL method to train the proposed Ptr-A* model in an unsupervised manner. Lastly, we propose three search strategies at inference.

We conduct comprehensive experiments to evaluate the performance of the proposed DRL algorithm. The simulation results show that the proposed DRL algorithm with different search strategies can produce better trajectories for the UAV when compared with the baseline techniques. In particular, DRL-active-10240 always produces the best results with different numbers of clusters of test instances. We also analyze the impact of different trained models on the results. The trained 40-clusters model is shown to be able to solve the trajectory planning problem of the UAV on large-scale clusters problems. The proposed DRL algorithm offers an appealing balance between performance and complexity. A key advantage of our proposed DRL algorithm is its generalization ability with respect to the number of clusters used for training. The model can be trained on small-scale clusters for faster training, and then can be used to solve larger-scale clusters problems. This makes it clearly more suitable for solving large-scale clusters problems as compared to the baseline techniques. 

As for future work, we are interested in exploring other search strategies at inference to further improve the performance. It would also be interesting to develop a distributed DRL algorithm based on the Ptr-A* model to solve multiple UAVs' trajectory planning problem jointly. We also plan to investigate and improve the generalization capability of the proposed DRL algorithm, i.e., when the number of nodes per cluster at inference is significantly larger than that used for training.

\section*{Acknowledgement}

This work was supported by an NSERC/Cisco Industrial Research Chair in Low-Power Wireless Access for Sensor Networks.

\appendices
      {\section{Example of Attention Mechanism \vspace{1.3\baselineskip}}\label{app:t1}
        Here, we give a detailed numerical example of 3-clusters network to explain how the attention mechanism works. In Fig.~\ref{figure2}, the input sequence $\bm{\mathcal{G}} = \{b_0, G_1, G_2, G_3\}$ is transformed into a sequence of latent memory states $\bm{\mathcal{E}} = \{e_0, e_1, e_2, e_3\}$, which is the input of the decoder network. At decoding step 0, we calculate correlations between all elements in $\bm{\mathcal{E}}$ and the start tag $v_{\text{go}}$ by (\ref{mu}) and (\ref{pro}), which can be expressed as
        \begin{align}
            \tag{A.1}
            u^0_0 &=
	        \varphi \tanh{\left(W_{1}e_0 + W_2h_0\right)} \\\tag{A.2}
	        u^0_1 &=
	        \varphi \tanh{\left(W_{1}e_1 + W_2h_0\right)} \\\tag{A.3}
	         u^0_2 &=
	        \varphi \tanh{\left(W_{1}e_2 + W_2h_0\right)} \\\tag{A.4}
	        u^0_3 &=
	        \varphi \tanh{\left(W_{1}e_3 + W_2h_0\right)}.
       \end{align}
      The learning parameters, namely $\varphi, W_{1}, W_{2}$, are initialized by the Xavier initialization method and trained by the Adam optimizer as explained in Section IV.A. Then, the \emph{softmax} function is used to normalize the vector $u^0=\{u^0_0, u^0_1, u^0_2, u^0_3\}$. For the sake of illustration, we assume that the four elements in $u^0$ determine the following four conditional probability values:
      \begin{align}
        &P(\pi_{0}=0|\bm{\mathcal{G}})\nonumber \\  \tag{A.5} &= \frac{\exp{\left(u^0_0\right)}}{\exp{\left(u^0_0\right)} + \exp{\left(u^0_1\right)} + \exp{\left(u^0_2\right)} + \exp{\left(u^0_3\right)}} = 0.6
      \end{align}
      \begin{align}
        &P(\pi_{0}=1|\bm{\mathcal{G}})\nonumber \\  \tag{A.6} &= \frac{\exp{\left(u^0_1\right)}}{\exp{\left(u^0_0\right)} + \exp{\left(u^0_1\right)} + \exp{\left(u^0_2\right)} + \exp{\left(u^0_3\right)}} = 0.1
      \end{align}
     \begin{align}
        &P(\pi_{0}=2|\bm{\mathcal{G}})\nonumber \\  \tag{A.7} &= \frac{\exp{\left(u^0_2\right)}}{\exp{\left(u^0_0\right)} + \exp{\left(u^0_1\right)} + \exp{\left(u^0_2\right)} + \exp{\left(u^0_3\right)}} = 0.1
      \end{align}
      \begin{align}
        &P(\pi_{0}=3|\bm{\mathcal{G}})\nonumber \\  \tag{A.8} &= \frac{\exp{\left(u^0_3\right)}}{\exp{\left(u^0_0\right)} + \exp{\left(u^0_1\right)} + \exp{\left(u^0_2\right)} + \exp{\left(u^0_3\right)}} = 0.2.
      \end{align}
      Since $e_0$ has the highest conditional probability value at decoding step 0, the output $\pi_0$ of this step points to the first element of $\bm{\mathcal{G}}$, $b_0$.
      At decoding step 1,  we use the same approach to calculate the correlations between $b_0$ and the remaining elements in $\bm{\mathcal{E}}$ as follows
      \begin{align}
            \tag{A.9}
	        u^1_1 &=
	        \varphi \tanh{\left(W_{1}e_1 + W_2h_1\right)} \\\tag{A.10}
	         u^1_2 &=
	        \varphi \tanh{\left(W_{1}e_2 + W_2h_1\right)} \\\tag{A.11}
	        u^1_3 &=
	        \varphi \tanh{\left(W_{1}e_3 + W_2h_1\right)}.
       \end{align}
      Similarly, by using the  \emph{softmax} function, we calculate the following conditional probabilities based on the obtained elements $\{u^1_1, u^1_2, u^1_3\}$:
      \begin{align}
        &P(\pi_{1}=1|\pi_0,\bm{\mathcal{G}})\nonumber \\  \tag{A.12} &= \frac{\exp{\left(u^1_1\right)}}{\exp{\left(u^1_2\right)} + \exp{\left(u^1_2\right)} + \exp{\left(u^1_3\right)}} = 0.2
      \end{align}
      \begin{align}
        &P(\pi_{1}=2|\pi_0,\bm{\mathcal{G}})\nonumber \\  \tag{A.13} &= \frac{\exp{\left(u^1_2\right)}}{\exp{\left(u^1_2\right)} + \exp{\left(u^1_2\right)} + \exp{\left(u^1_3\right)}} = 0.1
      \end{align}
      \begin{align}
        &P(\pi_{1}=3|\pi_0,\bm{\mathcal{G}})\nonumber \\  \tag{A.14} &= \frac{\exp{\left(u^1_3\right)}}{\exp{\left(u^1_2\right)} + \exp{\left(u^1_2\right)} + \exp{\left(u^1_3\right)}} = 0.7.
      \end{align}
      The output $\pi_1$ of this step points to $G_3$ because the conditional probability of $e_3$ is maximum. Then, the process repeats until we obtain the full output sequence of the decoder network as  $\{b_0, G_3, G_1, G_2\}$, i.e., $\{\pi_0, \pi_1, \pi_2, \pi_3\}$. }

\balance

\end{document}